\documentclass[twocolumn,preprintnumbers,amsmath,amssymb]{revtex4}
\usepackage{tabularx,graphicx}

\usepackage{color}
\usepackage{hyperref}

\hypersetup{
    colorlinks=true,
    linkcolor=blue,
    filecolor=blue,      
    urlcolor=blue,
}

\usepackage{color}

\newcommand{\bluee}{\textcolor{black}}

\usepackage{ulem}   

\begin{document}
 

\newcommand{\beq}{\begin{equation}}
\newcommand{\eeq}{\end{equation}}
\newcommand{\beqn}{\begin{eqnarray}}
\newcommand{\eeqn}{\end{eqnarray}}
\newcommand{\bmath}{\begin{subequations}}
\newcommand{\emath}{\end{subequations}}
\newcommand{\bra}[1]{\langle #1|}
\newcommand{\ket}[1]{|#1\rangle}

\title{On the  ``Author Correction: Magnetic field screening in hydrogen-rich high-temperature superconductors'',
 Nat Commun 14, 5322 (2023) }

\author{J. E. Hirsch}
\address{Department of Physics, University of California, San Diego,
La Jolla, CA 92093-0319}
 
 \begin{abstract} 
  I analyze 
   the implications of the recently published ``Author Correction''  (\href{https://www.nature.com/articles/s41467-023-40837-2}{Nat Commun 14, 5322 (2023)})
   to a paper by Eremets and coauthors reporting magnetization measurements on 
 hydrides under high pressure (\href{https://www.nature.com/articles/s41467-022-30782-x}{Nat Commun 13, 3194 (2022)})
 to the understanding of the validity and reproducibility of the published data.
 This paper is a compilation of several different papers already published or to be published in the scientific literature.
 \end{abstract}
 \maketitle 
 
 \section{Can linear transformations bend a straight line? }

 {\it 
 Ref. \cite{e2021p} claims to show magnetization measurements that demonstrate that sulfur hydride 
 and lanthanum hydride under
 pressure are  high temperature superconductors. In Ref. \cite{hmscreening}, it was pointed out that
 Figs. 3a and 3e of Ref. \cite{e2021p}  in its original form, Ref.  \cite{original}, were inconsistent with each other according to the figure caption and text in Ref. \cite{original}. 
 Recently, the authors of Ref. \cite{e2021p} published an ``Author Correction'' \cite{correction} to the original 
 form of Ref.
 \cite{e2021p} explaining 
  that several  linear transformations were used to obtain the data published in Fig. 3a from  measured
 data published in Fig. 3e. Here I show  that  data shown in Fig. 3a of Ref. \cite{e2021p} 
 are not related by $any$ set
 of linear transformations to  data published in Fig. 3e of Ref. \cite{e2021p}, contrary to
 what is stated in the ``Author Correction''  \cite{correction} and in the 
 corrected Ref. \cite{e2021p}. Implications of this finding are discussed.}

 \subsection{Introduction}

       \begin{figure} [t]
 \resizebox{8.5cm}{!}{\includegraphics[width=6cm]{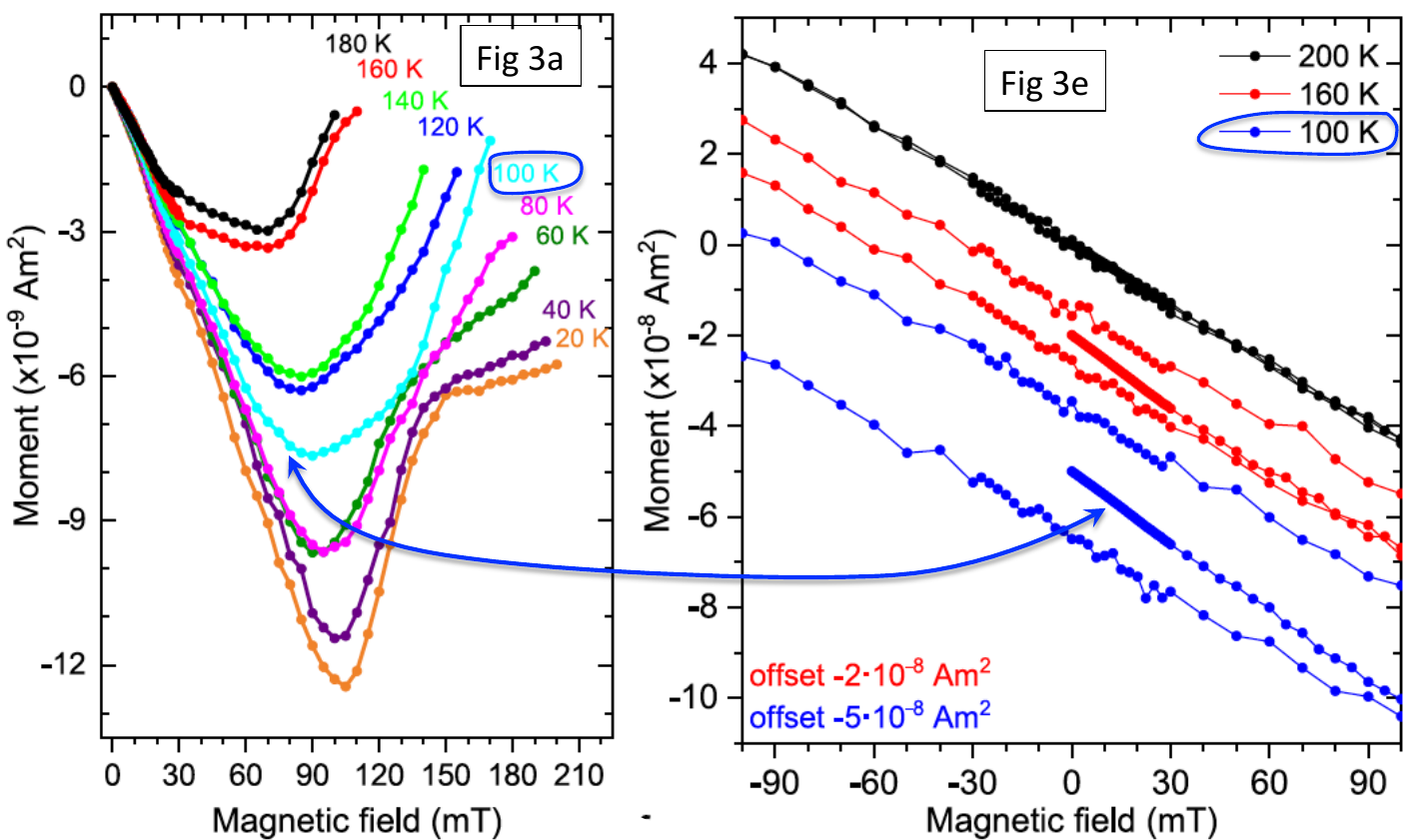}} 
 \caption { Fig. 3a (left panel) and Fig.  3e (right panel) of Ref. \cite{e2021p} showing induced magnetic moment versus magnetic
 field for a sample of $H_3S$ under high pressure.  We have added a two-sided arrow pointing to the
 curves for temperature 100 K in both panels that are the focus of this Comment, and have circled the labels indicating
 those curves.}
 \label{figure1}
 \end{figure}

 A linear transformation that leads from a set of numbers $\{x_i\}$ to another set of numbers $\{y_i\}$ is
 of the form
 \beq
 y_i=m_1x_i+b_1 .
 \eeq
 If we perform another linear transformation on the set of numbers $\{y_i\}$ to yield a 
 set of numbers $\{z_i\}$
  \beq
 z_i=m_2y_i+b_2 
 \eeq
 the transformation that relates $\{z_i\}$ to $\{x_i\}$ is
 \beq
 z_i=m_1m_2x_i+m_2b_1+b_2. \equiv m_3x_i+b_3
 \eeq
 and  is also a linear transformation. 
 
  If the set of points $\{x_i\}$ are results of measurements performed as a function of magnetic field with
 values  $H_i$ 
 and they are linearly related to $\{H_i\}$, they will lie on a straight line when plotted as function
 of $H_i$ and obey the relation
 \beq
 x_i=m_4H_i+b_4
 \eeq
 and if we perform any number of linear transformations on the measured data $\{x_i\}$ to obtain
 processed data $\{z_i\}$, the processed data will also be related to the magnetic field by a 
 linear relation
  \beq
 z_i=m_5H_i+b_5
 \eeq
 and lie on a straight line when plotted versus $H_i$.
 
 The left panel of Fig. 1 shows results for magnetic moment versus magnetic field
 for a sample of $H_3S$ under high pressure claimed to become superconducting
 at temperature $T_c=196K$, published as Fig. 3a in Ref.  \cite{e2021p}. 
 It was argued in Ref. \cite{e2021p} that the behavior shown in that figure shows that the
 sample is superconducting. Values for the lower critical magnetic field as
 function of temperature were obtained from the values of magnetic field where the curves on the left panel
 of Fig. 1 start to deviate from linearity (called $H_p(T)$ in Ref. \cite{e2021p}), those values were shown  as Fig. 3c in Ref. \cite{e2021p}.
 
 The right panel of Fig. 1  shows portions of hysteresis loops for the magnetic
 moment at
 three temperatures, published as Fig. 3e in Ref.  \cite{e2021p}.  Here we focus on the data for 100K, shown in blue. The middle curve is the
 virgin curve, obtained when the field is first turned on (the curves were shifted vertically for display), and the upper and lower blue curves are
 the hysteresis curves obtained when the field is lowered back to zero and negative values, and then
 increased again to positive values respectively. 
 
 The  figure caption of Fig. 3 and related text in Ref. \cite{e2021p} in the original form of Ref. 1 \cite{original}, 
 published June 9, 2022,  implied that the data in Fig. 3a were measured data. We pointed out in
 Ref. \cite{hmscreening} that the data shown in Fig. 3a and in the corresponding virgin curves of the Fig. 3e
 hysteresis loops (e.g. the middle blue and red curves in Fig. 3e)  should be identical. 
On September 1, 2023, a correction to Ref. \cite{e2021p}  was published, Ref. \cite{correction},
explaining that the original version \cite{original} was in error and  that the data shown in Fig. 3a were not the same as those in Fig. 3e because the former  were  obtained by performing a set of linear transformations on 
 the measured data of Fig. 3e. Ref. \cite{e2021p} was replaced with its corrected version on that date, which
 includes some information on these transformations. In the original form of the paper (Ref. \cite{original})
 there was no information that such transformations had been performed.

 We have recently discussed \cite{unpulling} the implications of the Author Correction \cite{correction}
  to the claims of Ref. \cite{e2021p} that $H_3S$ is a superconductor
 and argued that the inferred behavior of magnetic moment with magnetic field is in fact
 inconsistent with the conclusion that the system is a superconductor.
 
 In this Comment we focus on the relation between measured data and derived data, i.e. Fig. 3e and
 Fig. 3a of Ref. \cite{e2021p}, right and left panels of Fig. 1, for temperature T=100K. In Fig. 1 we have drawn a 
 two-sided arrow showing the data that we will be discussing. We will show that the data shown in
 Fig. 3a are not related to the data shown in Fig. 3e by linear transformations, contrary to what
 the authors state in Refs. \cite{correction} and \cite{e2021p}. The implications of this are then discussed.

          \begin{figure} [t]
 \resizebox{8.5cm}{!}{\includegraphics[width=6cm]{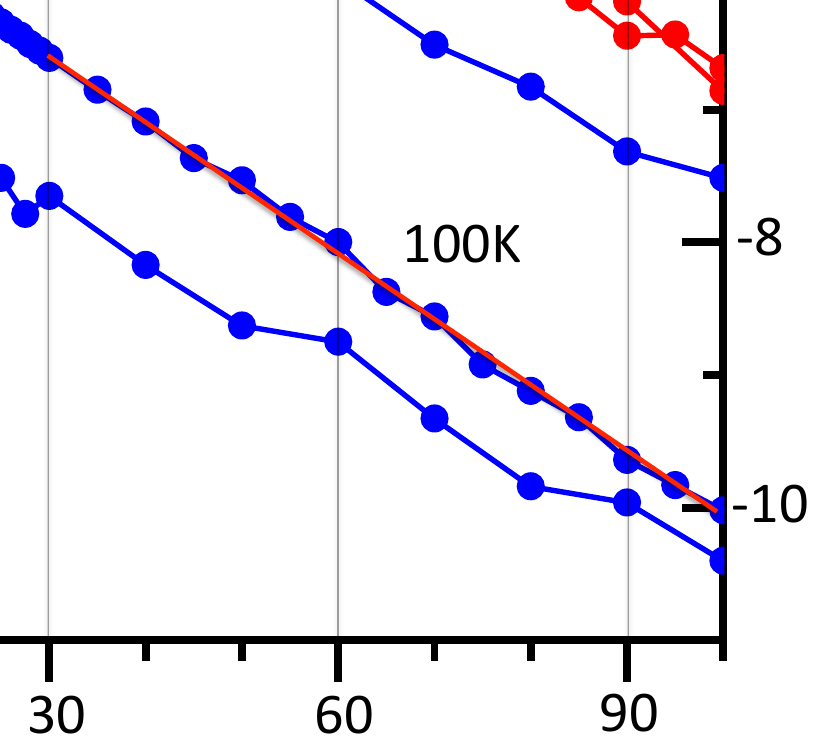}} 
 \caption { A portion of the right panel of Fig. 1, from Fig. 3e of Ref. \cite{e2021p}, showing the
 virgin curve for T=100K (middle blue curve). We have connected the points for magnetic
 field 30mT and 100mT by a straight red line. 
 The numbers on the horizontal axis give magnetic field values in mT and the numbers on the vertical
 axis give the magnetic moment in units $10^{-8} Am^2$.}
 \label{figure1}
 \end{figure}

          \begin{figure} [t]
 \resizebox{7.5cm}{!}{\includegraphics[width=6cm]{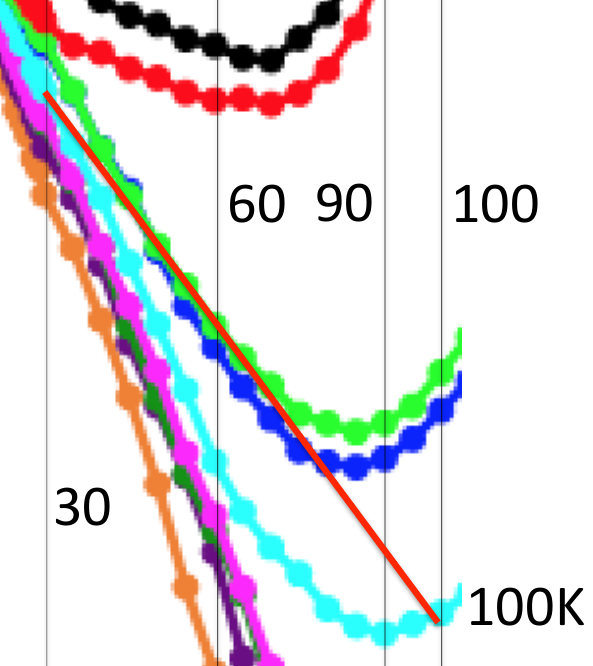}} 
 \caption { A portion of the left panel of Fig. 1, from Fig. 3a of Ref. \cite{e2021p}.
 We added the vertical thin lines, the numbers next to the   lines give the values of magnetic field
 for those points, in mT. 
 The light blue dots connected with a light blue line show the magnetic moment data for temperature 100K. Presumably, the dots are data and the connecting line was drawn to guide the eye.
 We have added a straight red line connecting the points for 30mT and 100 mT. Note that
 the data for field values between 30 mT and 100 mT lie all below the red line and do not follow a
 linear behavior.}
 \label{figure1}
 \end{figure}

        \begin{figure} [t]
 \resizebox{7.5cm}{!}{\includegraphics[width=6cm]{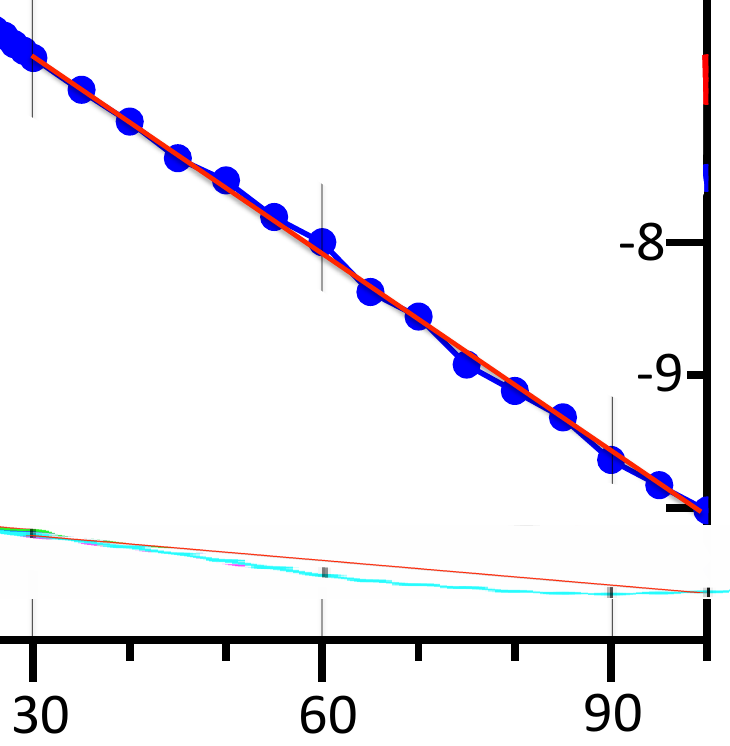}} 
 \caption {Comparison of the 100K data for the virgin curve in Fig. 3e of Ref. \cite{e2021p}
  (upper blue points) and 100K data of Fig. 3a of Ref. \cite{e2021p} (lower light
 blue points) on the same scale. The red lines connect the points at field 30mT and 100mT
 in both cases. 
 The numbers on the horizontal axis give magnetic field values in mT and the numbers on the vertical
 axis give the magnetic moment in units $10^{-8} Am^2$. }
 \label{figure1}
 \end{figure}

 \subsection{Analysis}
 
 A portion of the data shown on the right panel of Fig. 1, including part of the data for the three blue curves corresponding to temperature T=100K, is shown in Fig. 2. 
 We have added a straight red line connecting the points for magnetic field values 
 30 mT and 100 mT of the middle  blue curve which is the virgin curve. It can be seen that all the points lie approximately on this straight line, with
 no systematic deviation from it.

In Fig. 3 we show a portion of the data shown on the left panel of Fig. 1, containing the 
100K points (in light blue) for magnetic field between 30mT and 100 mT. 
We have drawn a straight red line connecting the light blue  points for 30 mT and 100 mT.
It can be seen that there is a systematic deviation of the light blue points from this red line.
All light blue points, except for the first and last points, lie below the straight red line.
In other words, the data in Fig. 3 are not linearly related to the magnetic field in this range of fields.
 
 The vertical scales in Fig. 2 and Fig. 3 are rather different, so to make a more direct comparison
 we plot in Fig. 4  the data of 
 Fig. 2 and Fig. 3 with the same horizontal and vertical scales.
 The former are clearly approximately linear in field, the latter are clearly not.
 This implies that the latter have not been obtained from the former by a set of linear
 transformations.

 \subsection{Implications}
 
 The ``Author Correction'' \cite{correction} states, in explaining the relation of the data in Fig. 3e to the
 data in Fig. 3a of Ref.  \cite{e2021p}, what is now the figure caption of Fig. 3 of Ref. \cite{e2021p}:
 
 {\it ``a, b Magnetic moment associated with the penetration of the applied magnetic field into the Im-3m-H3S  phase at P=155  $\pm$ 5 GPa and the Fm-3m-LaH10    phase at  P=130 $\pm$ 8 GPa  based on virgin curves of the 
 M(H) magnetization data at selected temperatures. The curves were superimposed by performing linear transformations for a better representation. A linear background, defined as a straight line connecting 
 M(H=0T) and M(H=1T)  at corresponding temperature, was subtracted. After that the data were normalized to 
 H= 15 mT  data so that to have the same initial linear M(H)  slope.''  }
 
\noindent and in the text associated with Fig. 3  it explains how the value of the critical field shown in 
 Fig. 3c of the paper was obtained:
 
 {\it ``The value of $H_p$, at which an applied magnetic field starts to penetrate into the sample, was determined from the onset of the evident deviation of the M(H)  from the linear dependence.''}
 
\noindent and in explaining how the data in Fig. 3a were obtained it states 
 
{\it ``To better illustrate the determination of $H_p$ in Figs. 3a and 3b, we have subtracted a linear background from the measured M(H)  magnetization data. This linear background was determined as the straight line connecting two endpoints: the magnetic moment value at H=0 T    (the starting point of measurements) and the magnetic moment value at H=  1 T (the highest value of the applied magnetic field) (see Supplementary Fig. S12). Subsequently, we performed additional linear transformations so that the curves have the same initial linear M(H) slope. Importantly, these linear manipulations do not affect the onset of the deviation of the M(H)  virgin curve from the linear dependence.''} 
 
Note that the subtraction of a linear background is also a linear transformation.
 It is clear that the determination of $H_p$ hinges on the crucial fact that all these transformations and
 {\it ``linear manipulations''} performed were linear, since $H_p$ was determined by the deviation
 of the transformed data from linearity. If any of the transformations performed in going from the
 measured data to the data from which $H_p$ was inferred was non-linear, the procedure would
 clearly be invalid. 
 
We have shown here that not all transformations performed could have been linear,
 since from the linear data in Fig. 2, non-linear data in Fig. 3 were derived by these transformations. 
  This renders the determination of $H_p$, a key conclusion of  Ref. \cite{e2021p}, invalid.

\subsection{Conclusions}
 In order to have confidence that Figs. 3a and 3b of Ref. \cite{e2021p}, claimed to be evidence
 that $H_3S$ and $LaH_{10}$ under pressure are high temperature superconductors \cite{e2021p},
 reflect the real physics of the samples under study,
 it is important to understand the relation between the data shown in those figures and
 the measured data. We have shown here that for temperature $T=100K$ the data in Fig. 3a were
 not obtained by linear transformations from measured data, contrary to what is claimed by the authors 
 \cite{e2021p,correction}.
 We have also shown in Ref. \cite{unpulling} that the data for $T=160K$ in Fig. 3a were
 not obtained from the raw data for that temperature shown in Fig. S12c of Ref. \cite{e2021p},
 as was  claimed in the paper \cite{e2021p} and its Author Correction \cite{correction}. 
 
 
 It cannot be ruled out, given the available information, that several different experimental runs were performed as a function of magnetic field
 for the same temperature and pressure, and that the data shown in Fig. 3a were derived through linear transformations performed on measured data from a different run than the measured data shown in
 Fig. 3e. However, if this was the reality it   would raise several additional questions requiring
 answers, namely: (i) why would one 
 experimental run show a systematic deviation from linearity, and a different run under the same conditions show no deviation from linearity? (ii) If 
 different runs showed different   behavior, what was the criterion used by the authors to decide than one run (the one showing deviation from
 linearity) should be kept and the other run (not showing deviation from linearity) should be  ignored in order to determine the
 value of $H_p$? (iii) Why didn't the authors discuss this issue, neither in the original version of the paper \cite{original} 
 nor in the Author Correction \cite{correction} nor in the corrected version of the paper \cite{e2021p}?
 
 Another important unanswered question is, why are the data shown  in Fig. 3e only for temperatures 100K and above, and there is no information in Ref. \cite{e2021p} on the measured data that gave rise to the four curves at
 temperatures lower than 100K shown in Fig. 3a? For lower temperatures the non-linear temperature dependence of the magnetic moments shown in Fig. 3a in the range of fields from 0 to 200 mT is very prominent. Thus it should be easily discernable in curves such as shown in Fig. 3e for lower temperatures.
 
 Having information on the measured data from which the data shown in Fig. 3a (and 3b)  were derived would allow readers to come to their own conclusions on 
 what the implications of the data shown in Fig. 3a (and 3b)  are to the understanding of the physical properties
 of the samples under study. Not having that information, nor detailed information on the transformations
 and manipulations that
 were performed on the measured data to arrive at the data published in Fig. 3a and 3b, prevents readers from having 
 confidence that the data shown in Fig. 3a and 3b reflect the physical properties of the samples under study. 
 We have asked
 the authors of Ref. \cite{e2021p} to release the measured data underlying the published results for all the
 curves shown in Figs. 3a and 3b,
 in accordance with the ``Data availability'' statement published with the paper \cite{e2021p}. The data have
 not been released to date.
 
 Ref. \cite{e2021p} is the strongest magnetic 
 evidence presented to date that hydrides under pressure are high temperature superconductors \cite{pers}.
 Quoting from Ref. \cite{e2021p}, for hydride materials under pressure
 {\it ``Magnetic measurements, which are, inter alia, a crucial and independent
test of superconductivity are scarce.''} The only other magnetization measurements for hydrides under pressure 
  in the literature are  in the first publication on sulfur hydride back in 2015 \cite{h3s}, about which Ref. \cite{e2021p} says 
 {\it ``$H_{c1}$ was only roughly estimated from the hysteretic loops of M(H)
data instead of the initial virgin portion of magnetization curves of
zero-field-cooled (ZFC) sample''} as is done in Ref. \cite{e2021p}. Thus, it is very important to the understanding
 of hydrides under pressure claimed to be high temperature superconductors to understand the significance of the results published in Ref. \cite{e2021p}  discussed in this Comment.

  Given this situation, we argue that it is imperative that the authors of Ref. \cite{e2021p}  (i) release the underlying measured data, and
  (ii) provide an explanation for how the  data plotted in Fig. 3a were derived from the measured data.
   The explanation should be either a Reply to this Comment or an ``Author Correction'' to the 
  ``Author Correction''\cite{correction}. With  those two pieces  of information, readers will have the needed
  elements to decide to what extent in their judgement  the data shown in Ref. \cite{e2021p}  contribute to the
  understanding of the physics of these materials. Absent  that information, I argue that in view of the
  data analysis presented in this Comment     the conclusions of Ref. \cite{e2021p} derived from  its  published data
cannot be interpreted as being supported by the experimental data.

 \section{Matters Arising on the Author Correction to ``Magnetic field screening in hydride superconductors''}

 {\it Ref. \cite{e2021p} reported measurements of diamagnetic moment versus magnetic field for sulfur hydride 
 and lanthanum hydride under
 pressure in its Figs. 3a and 3b respectively, claiming that they provide evidence that the samples are superconducting and allow to infer
 the value of the lower critical magnetic fields as function of temperature.
 Ref. \cite{correction} explained that several $linear$ transformations were used in obtaining the data shown
 in Figs. 3a and 3b of Ref. \cite{e2021p} from the measured data shown in Figs. 3e and S10, and 3f and
 S11 of Ref. \cite{e2021p} respectively, including subtraction of a significant diamagnetic background. Here we show that 
 those statements are contradicted by facts. This \bluee{calls into question} the claim of Ref. \cite{e2021p} that Figs. 3a and 3b of Ref. \cite{e2021p}  are
 evidence for superconductivity in these materials.  }
  \newline 
 
 From the deviation of linear dependence of magnetic moment on magnetic field for 
 $LaH_{10}$ under pressure shown in Fig. 5, reproduced from Fig. 3b of Ref. \cite{e2021p}, the authors extracted values of critical field versus temperature.
 As the Author Correction to Ref. \cite{e2021p} recently published \cite{correction} explains, the data
 shown in Fig. 5 are not measured data. Rather, they were obtained from measured data by a set
 of linear transformations. Ref. \cite{correction} explains that linear transformations would
 {\it ``not affect the onset of the
deviation of the M(H) virgin curve from the linear dependence''} on magnetic field.
 
 The latter statement is correct. It necessitates that the transformation is truly linear, if the transformation
 is not linear the statement is clearly invalid. 
 For the virgin curve, one point is fixed at the origin, since there is no magnetization for zero
 applied field. In Fig. 5 we have connected the origin and the point for the T=80K curve for 
 $LaH_{10}$ at field H=100 mT by a straight red line. Note that all the points for the magnetic 
 moment for T=80K for fields between 0 and 100 mT fall below the straight red line.
 
 The data for Fig. 5 were obtained from measured data shown in Fig. 3f of Ref. \cite{e2021p}. In Fig. 6 we show
 a portion of those data. The three blue curves correspond to T=80K, as the blue points in Fig. 5. 
 The middle blue curve is the virgin curve, starting with zero moment at zero field.
 We have connected that point with the moment at field H=100 mT by a straight red line.
 
 It can be seen in Fig. 6 that some of the measured points fall below the straight red line and some fall above.
 This is different from the blue points in Fig. 5, that are all below the straight red line connecting the points at 
 zero field and 100 mT field.
 
         \begin{figure} [t]
 \resizebox{8.5cm}{!}{\includegraphics[width=6cm]{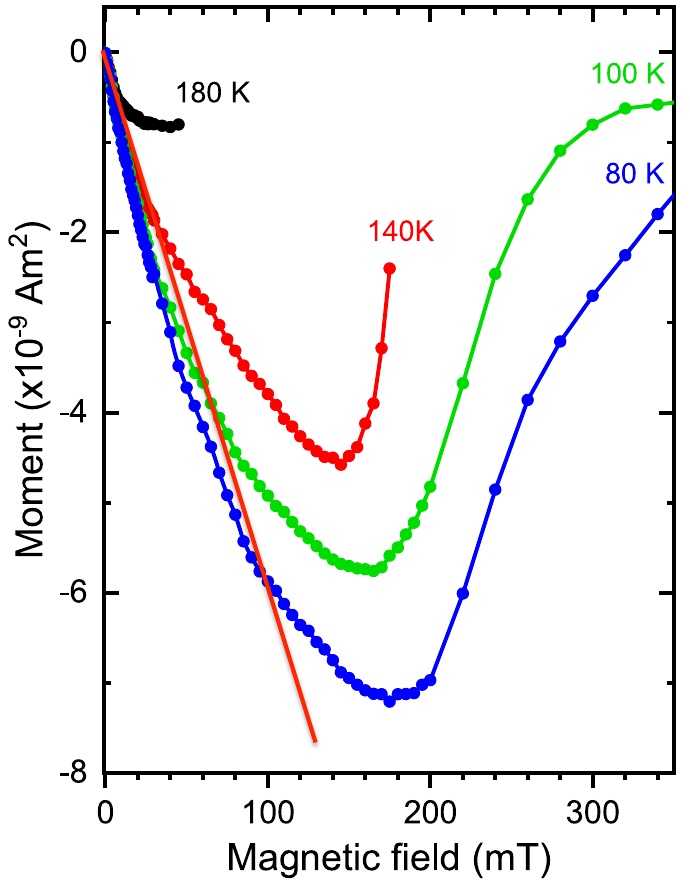}} 
 \caption {Magnetic moment of $LaH_{10}$ under pressure versus magnetic field, from
 Fig. 3b of Ref. \cite{e2021p}. We have added the straight red line connecting the origin with the
 magnetic moment at field 100 mT for the curve for temperature 80K (blue points). }
 \label{figure1}
 \end{figure} 
 
         \begin{figure} [t]
 \resizebox{8.5cm}{!}{\includegraphics[width=6cm]{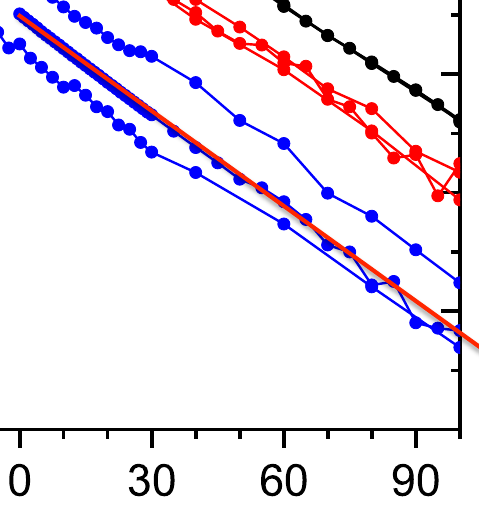}} 
 \caption {Magnetic moment of $LaH_{10}$ under pressure versus magnetic field, from
 Fig. 3f of Ref. \cite{e2021p}. The center blue curve is the virgin curve for temperature 80K. We have added a straight red line connecting the origin with the
 magnetic moment at field 100 mT for the virgin curve for temperature 80K (blue points). 
 The numbers on the bottom axis give the magnetic field in mT.}
 \label{figure1}
 \end{figure} 
 
          \begin{figure} [t]
 \resizebox{8.5cm}{!}{\includegraphics[width=6cm]{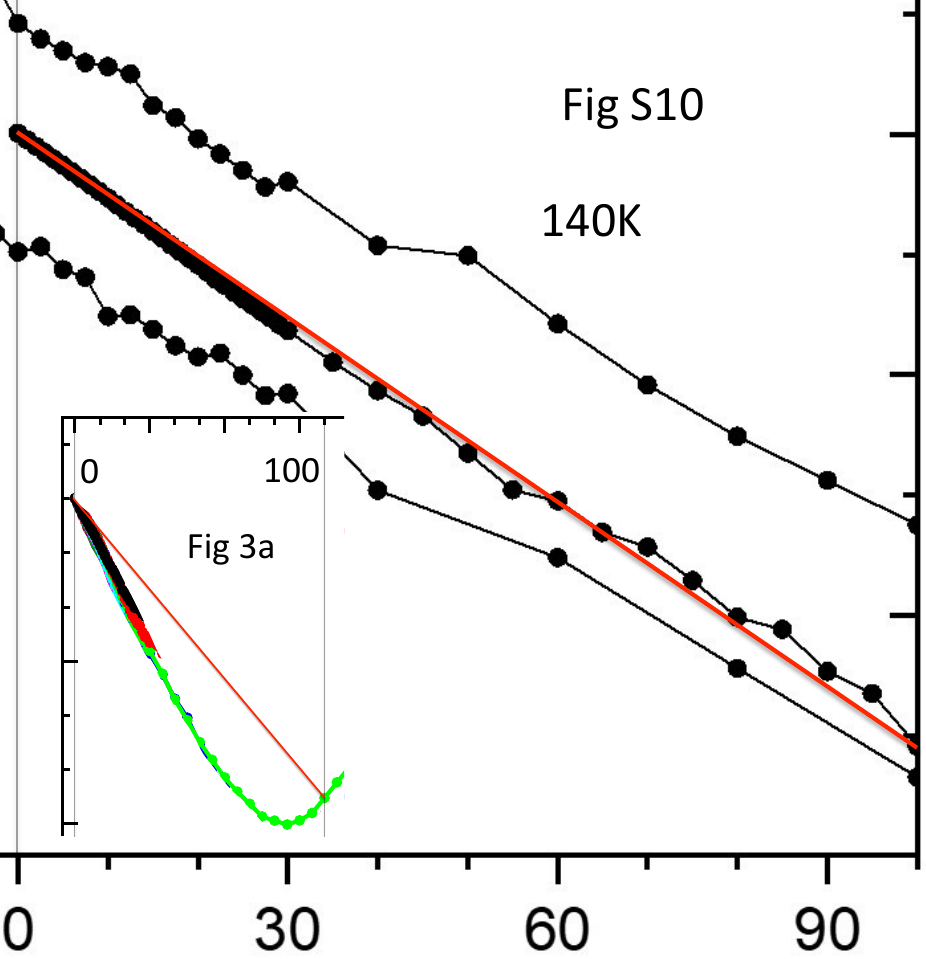}} 
 \caption {Magnetic moment of $H_3S$ under pressure versus magnetic field for temperature 140K, from
 Fig. S10 of Ref. \cite{e2021p}. The center black curve is the virgin curve. We have added a straight red line connecting the origin with the
 magnetic moment at field 100 mT.  The numbers on the bottom axis give the magnetic field in mT.
 The inset at the lower left shows the magnetic moment from Fig. 3a of Ref. \cite{e2021p} 
 at 140K supposedly obtained after 
 linear transformations, the red line in the inset connects the points at field 0 and 100 mT.
}
 \label{figure1}
 \end{figure} 

 Given any linear transformation of the data points in Fig. 6, we can redraw the red line so that
 it passes through the transformed points at 0 mT and 100 mT, and again some of the data points will fall below
 the new straight line and some above. 
Thus, the blue points in Fig. 5 could not have originated from a linear transformation nor from 
 a set of linear transformations applied
 to the blue points in Fig. 6, as Ref. \cite{correction} claimed. If the transformation used
 to obtain the curve shown in Fig. 5 from the curve shown in Fig. 6 was nonlinear, the procedure
 used to extract the critical field from the data in Fig. 5 is clearly invalid.
 
 The same is true for magnetization data of $H_3S$ shown in Fig. 3a of Ref. \cite{e2021p},
 supposedly derived through linear transformations of the data shown in Figs. 3e and S10
 of Ref. \cite{e2021p}. 
In Fig. 7 we show data for the hysteresis cycle for $H_3S$ at 140 K, from Fig.
 S10 of Ref. \cite{e2021p}. We have connected the points for zero magnetic field and 100 mT with a 
 straight red line. It can be seen that about half the points for fields between 30 mT and 100 mT are above the red line and half are
 below. Instead, the lower left inset in Fig. 7 shows the data derived from these measurements through supposedly
 linear transformations, reproduced from Fig. 3a of Ref. \cite{e2021p}. It can be seen that all the  green points in the
 curve lie below the straight red line connecting the points for 0 and 100 mT. It is obvious that those 
 points cannot results from linear transformations performed on the data shown in the 
 main body of the figure. 
 \bluee{Furthermore, no smoothing of the raw data shown in the main body of Fig. 7 can give rise to
 the parabolic green curve shown in the inset of Fig. 7.}
 \bluee{As shown in Ref.  \cite{bending}},
 the same applies to the magnetization data for $H_3S$ at temperature
 T=100K shown in Figs. 3a and 3e of Ref. \cite{e2021p}, 
 
 Therefore, the origin of the curves in Figs. 3a and 3b of Ref. \cite{e2021p} is not 
 what the paper and its Author Correction  states it is. What the origin  of those curves is is 
 unknown.  \bluee{Furthermore, the origin of the {\it data points (circles)}
 on the curves in Figs. 3a and 3b and how they were derived from the measured data
 points is unknown.} If the
 transformations used to obtain them from measured data were nonlinear, as the analysis in this paper
  \bluee{indicates}, 
 it would not be possible to extract values of critical field from deviations from linearity, as
 Ref. \cite{e2021p} does. More generally,  the data shown
 in Fig. 3a and 3b cannot be interpreted as showing physical properties of the samples, since
 their relation to the measured data is unknown. They certainly  cannot be used to infer the values of
 the critical field shown in Figs. 3c and 3d of Ref. \cite{e2021p}, a central result of
 Ref. \cite{e2021p}, unless the relation of the   curves   shown in Figs. 3a and 3b with  the measured data can be clearly established.
  \bluee{In particular, for the lower temperature curves of Fig. 3 of Ref. \cite{e2021p}, T=20K, 40K, 60K, 80K, crucial to obtain the
 zero-temperature value of the lower critical field of $H_3S$ deduced in the paper, there is no information in Ref. \cite{e2021p} nor in Ref. \cite{correction} about
 the measured data from which the published data were derived.}

 To clarify the relation between the   \bluee{information conveyed} in Figs. 3a and 3b and the measured data, the authors 
 of Ref. \cite{e2021p} should 
 release their measured data for examination by readers. We have requested access to those data on
 January 11, 2023, and repeatedly therafter. The data have not been released \bluee{to date}.

  Discussion of other aspects of the magnetization data of Ref. \cite{e2021p} and its Author
  Correction \cite{correction} and their implications for the question whether or not they provide evidence for
  superconductivity in  these materials  is given in Ref\bluee{s}. \bluee{\cite{bending,unpulling,hysteresis}}.

 \section{Hysteresis loops  in  measurements of  the magnetic moment of hydrides under high pressure: implications
for superconductivity}

{\it Measurements of  magnetic moments of hydride materials under high pressure have been claimed to prove the existence of superconductivity
 in these materials \cite{e2015, e2021,e2021p,correction,e2022p,etrapped}. However, detection of the signal from the small sample requires subtraction of a large background contribution
 whose details are largely unknown. Here we analyze reported measurements and point out that 
 the resulting hysteresis loops are incompatible with the conclusion that they result from superconductivity,
 independent of what assumptions are made about the background signal.  We argue that this also invalidates 
 the conclusion that the magnetic moment  measured  after the external magnetic field is turned off is
 evidence for trapped magnetic flux resulting from superconducting currents, as proposed in Ref.  \cite{etrapped}. 
Our results imply that to date no magnetic evidence for
 the existence of high temperature superconductivity in hydrides under pressure exists, despite multiple claims to the contrary.}

 \subsection{Introduction}
 In 2014 \cite{e2014} and 2015 \cite{e2015}, A.P. Drozdov, Mikhail Eremets and coworkers reported the discovery of high temperature superconductivity in
 sulfur hydride under high pressure, confirming theoretical predictions \cite{h2spred,h3spred} and expectations \cite{ashcroft2} based on conventional BCS-electron-phonon theory \cite{pickett}.
 This launched the development of a new highly active area of research. By now, more than two dozen hydrogen rich materials have been claimed to be high temperature superconductors based on experimental evidence \cite{troyan}, and hundreds  such materials have been claimed to be high temperature superconductors based on theoretical evidence \cite{zurek}.
 
 The experimental claims are mostly based on resistance measurements. However, as acknowledged by the creator of the field himself
 \cite{ehistory}, {\it ``To guarantee that superconductivity occurs, the most crucial signature that must be
present is the expulsion of the magnetic field below $T_c$ (the Meissner effect). Without these
measurements, Science and Nature rejected our paper even if we had already substantial
shreds of evidence of superconductivity''}. 
 
Nature did accept the 2015 paper \cite{e2015} even though it presented   \bluee{no}  evidence for {\it ``expulsion of the magnetic field
 below $T_c$''}. In fact, not a shred of evidence for expulsion of magnetic fields in these materials exists even today \cite{hmscreening}. 
 But the 2015 paper did present results of  magnetic moment measurements using a SQUID magnetometer  that appeared to show
   the development of a diamagnetic moment upon application of
 a magnetic field at low temperatures, and that was deemed sufficient \cite{edwin} for publication in Nature at that time \cite{e2015}.

 No new measurements of magnetic moments of these materials were reported for the ensuing 6 years.
 Then in 2021 and 2022, new measurements of magnetic moment of sulfur hydride again  using a SQUID magnetometer were reported   by 
 V. S. Minkov, Mikhail Eremets and
 coworkers \cite{e2021,e2021p,correction,e2022p,etrapped}. These measurements are the subject of this paper.
 No other magnetic moment  measurements on these materials  by other groups using SQUIDs have been reported to date.

 \subsection{Magnetic measurements in 2022 vs in 2015}
 
  \bluee{It is reasonable to expect} that with the passage of time and improvements in the experimental  equipment, sample preparation methods and
 experimental techniques, magnetic measurements
 should become more accurate and indicative of the  physics of the samples  under study. Indeed that
 seems to be the view of the authors of these papers \cite{e2015, e2021,e2021p,correction,e2022p,etrapped}, as illustrated in the following.
 
 In Ref. \cite{e2022p} of 2022, the authors state that the superconducting transition detected through magnetic measurements in the
 2015 work {\it ``fully agrees`with the recent and more accurate 
 measurements \cite{e2021}''} of 2021.  Furthermore, the authors in 2022 \cite{e2022p} state that
 {\it ``Only recently, we succeeded in improving our measurements
of the magnetic susceptibility significantly \cite{e2021} (see
Fig. 5c-h). For that we used a solid $BH_3NH_3$
as an alternative
source of hydrogen instead of pure hydrogen. This
allowed us to produce large samples with a diameter close to
that of the culet of the diamond anvils, and to obtain a pronounced
diamagnetic signal from superconducting phases
under high pressures''}.

 We start by comparing hysteresis loops without background subtraction measured in 2015 \cite{e2022p} and 2022 \cite{e2021p}, shown in 
 Fig. 8 at temperature T=100K and similar pressures (140 GPa and 155 GPa respectively). It is surprising that the background was paramagnetic in 2015 and is diamagnetic in 2022, as indicated by the sign of the slopes of the 
 curves  in Fig. 8,
 given that for both experiments the cell used was reported to be  a non-magnetic cell
 made of Cu-Ti alloy \cite{e2015,e2021p} 
 {\it ``in order to minimize the magnetic signal over a wide temperature range \cite{e2021p}''}.
 It is furthermore surprising that the background signal is   four  times $larger$ in 2022 compared to 2015, given that
 it would be desirable to design the experiment so as to minimize the magnitude of the background signal that
 can mask the behavior of the sample under study.

 Fig. 9 shows the hysteresis loop obtained from the 2015 measurements after   subtraction of a background signal 
 measured at 210K \cite{e2022p}.  The behavior 
 shown resembles what is seen in type II superconductors. 
 
            \begin{figure} [t]
 \resizebox{8.5cm}{!}{\includegraphics[width=6cm]{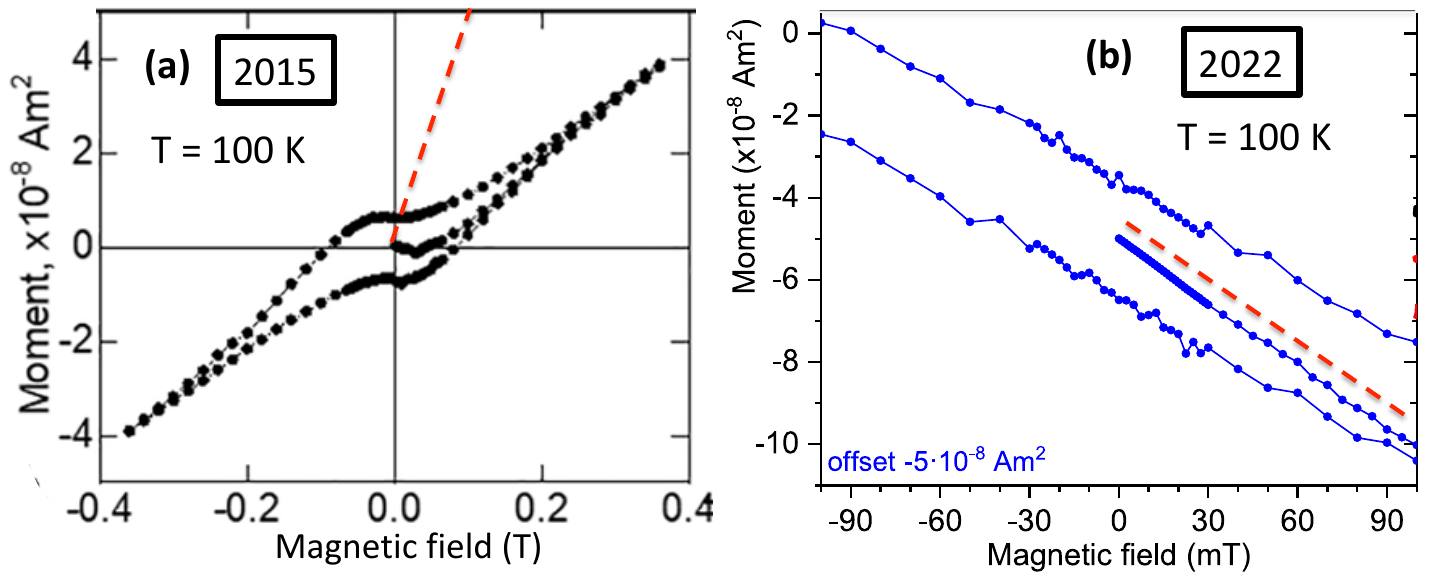}} 
 \caption {Comparison of magnetic moment versus magnetic field at 100K measured in 2015  and 2022, without
 background subtraction,
 as reported in Refs. \cite{e2022p} and \cite{e2021p} respectively. The background is paramagnetic in 2015 and diamagnetic in 2022.
 The slopes of the curves are approximately 4 times larger in 2022 compared to 2015, as indicated by the 
 dashed red lines that  we added that have the same slope in both panels.}
 \label{figure1}
 \end{figure} 
 
          \begin{figure} [t]
 \resizebox{8.5cm}{!}{\includegraphics[width=6cm]{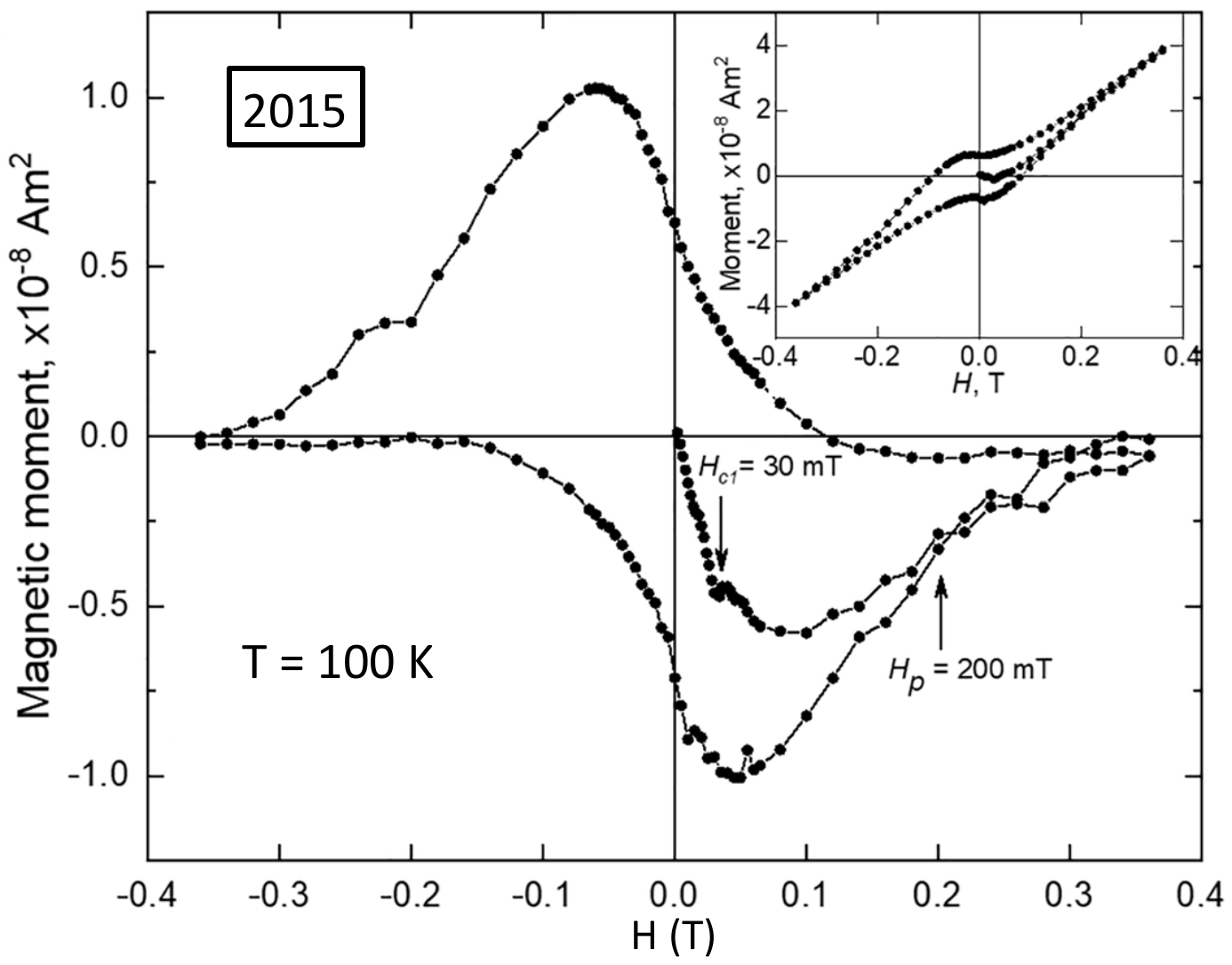}} 
 \caption {Magnetic moment of sulfur hydride versus magnetic field measured in 2015, reported in 2022 \cite{e2022p}.
 The figure caption reads 
 {\it ``M(H) data measured at
100 K showing the initial virgin curve and hysteretic loop on alternating
the magnetic field. Inset shows the original hysteretic loop before
subtraction of the background signal originating from the DAC at
210 K''}. }
 \label{figure1}
 \end{figure} 
 
           \begin{figure} [t]
 \resizebox{8.5cm}{!}{\includegraphics[width=6cm]{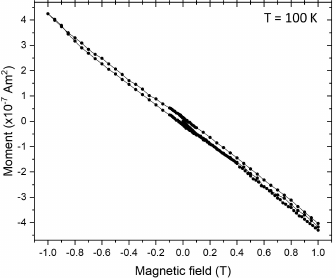}} 
 \caption {Hysteresis loop for the magnetic moment of sulfur hydride versus magnetic field measured in 2022, reported in
 Fig. S10 of  Ref.  \cite{e2021p},
 for magnetic fields between -1T and 1T.
 }
 \label{figure1}
 \end{figure} 
  
 Fig. 10 shows the hysteresis loop measured in 2022 \cite{e2021p} for field between -1T and 1T at 100K. We would like to analyze its behavior
 by looking at the numerical values of the data. Unfortunately, the authors have declined to make those data available despite
 repeated requests and despite the paper's stated Data Availability statement \cite{hmscreening,unpulling}. 
 We discuss our analysis based on the published figures in the next sections.
 
 \subsection{Analysis of extraction of lower critical field}

  Before analyzing the hysteresis loops we would like to discuss the extraction of lower critical field.  In Ref. \cite{e2022p} of 2022, the authors state, referring to the 2015 work Ref. \cite{e2015} {\it ``A lower critical magnetic field $H_{c1}$  of $\sim$ 30 mT was only
roughly estimated as the point associated to the bending of
the M(H) hysteretic loop''}, and that instead {\it ``the correct value of the inflection point in the
M(H) virgin curves was determined in the recent work 
and its value is higher $\sim$ 96 mT at 0 K for $H_3S$''}, with ``recent work'' referring to Ref. \cite{e2021p}.
 
 Fig. 11 shows on the left panel the magnetic moment versus field virgin curves reported in Fig. 3a of Ref. \cite{e2021p}.
 From the point where the curves start to deviate from linearity, the value of the lower critical field was extracted,
 which was then plotted in Fig. 3c of Ref. \cite{e2021p} versus temperature. That plot is shown on the inset of the right
 panel of Fig. 11.
 
 Focusing on temperature T=100K, the light blue curve on the left panel of Fig. 11 starts to deviate from linearity around
 H=67 mT, and the corresponding point for the critical field versus temperature in the inset of Fig. 11 (from Fig. 3c of
 Ref. \cite{e2021p}) is indicated by an arrow. The curve in the inset reaches its maximum  value 96mT at zero temperature.

 However, we have recently pointed out that in fact the measured data shown in Fig. 3e of Ref. \cite{e2021p} (Fig. 8 right panel here) cannot give rise to the 
 100K curve shown on the left panel of Fig. 11 \cite{bending}. On the right panel of Fig. 11 we show what we obtained from
 digitizing the measured  data
 shown on the right panel of Fig. 8 and subtracting  a linear background, obtained by connecting the H=0 and H=1T values
 of magnetic moment shown in Fig. 10, which is the prescription given in Ref. \cite{correction} to subtract the background.
 
 It can be seen in Fig. 11 that the points on the right panel and the light blue curve on the left panel look rather different. In particular, it is
 \bluee{not possible}  to decide from the data on the right panel that a deviation from linearity sets in at 
 H=67 mT, while it is plausible to do so from the data on the left panel.  We don't know the detailed behavior of the measured data immediately above 100 mT (the curves shown in Fig. 10 have very
 low resolution), but it would appear from the right panel of Fig. 11 that the data derived from the measured data would also be consistent with a 
 lower critical field of 100mT or larger, which would be larger than the zero temperature value shown
 in the inset.
 
 It is also apparent from Fig. 11 that the data on the right panel, that we extracted from the measured data shown in Fig. 
 3e of Ref. \cite{e2021p},
have a lot more scatter than the data reported in Fig. 3a of Ref. \cite{e2021p} shown on the left panel.
This indicates that, contrary to a recent suggestion \cite{reviewers}, it is unlikely that the data shown in Fig. 3a of Ref. \cite{e2021p} originated from a different measurement run of similar nature to   the measurements performed to obtain the reported measured data
in Fig. 3e of Ref. \cite{e2021p}.

            \begin{figure} [t]
 \resizebox{8.5cm}{!}{\includegraphics[width=6cm]{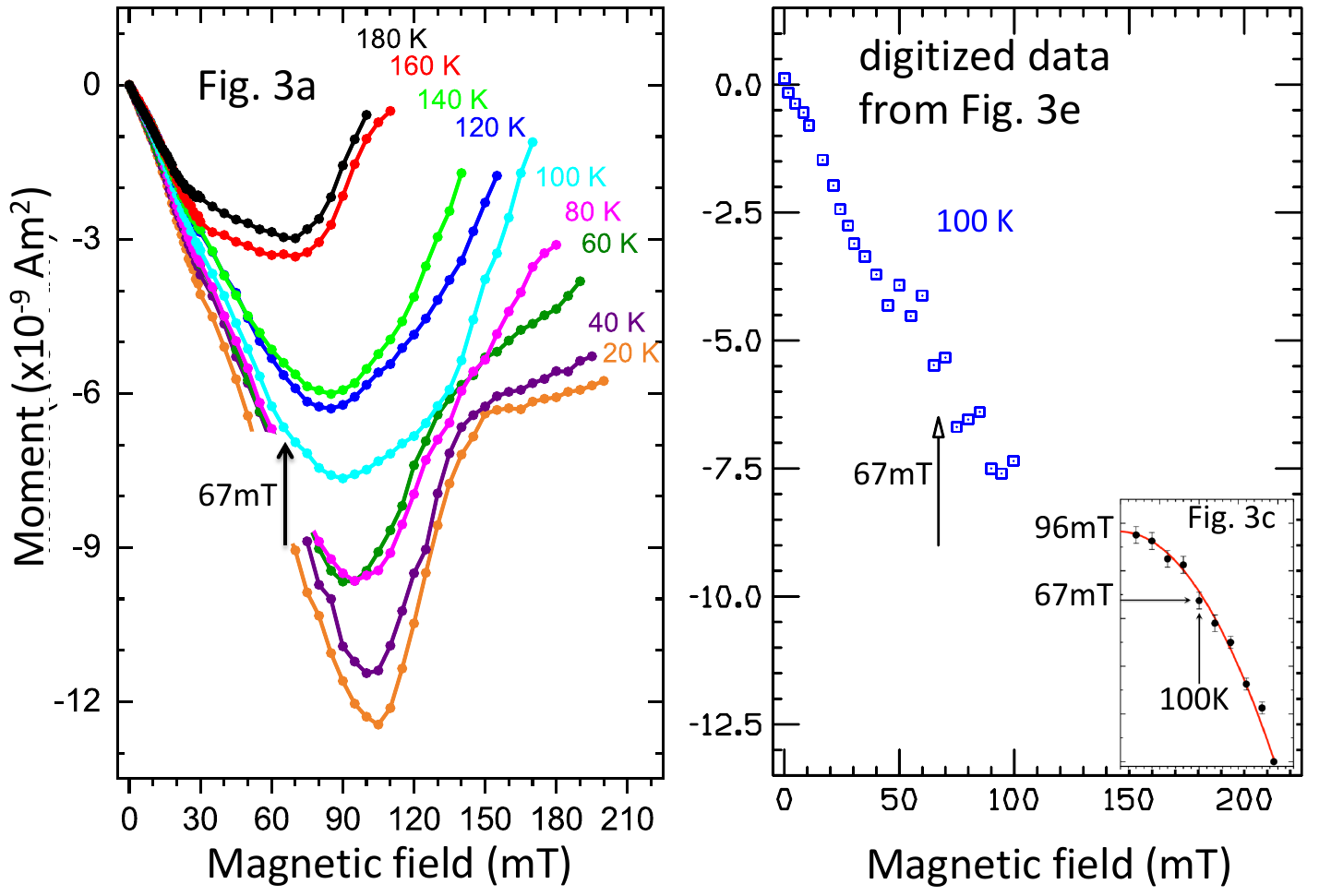}} 
 \caption { Left panel: magnetic moment versus field reported in Fig. 3a of Ref. \cite{e2021p}, 
 reportedly obtained by subtraction of a linear background from the measured data reported in 
 Fig. 3e of Ref. \cite{e2021p}. Right panel: results obtained by us by 
 subtraction of a linear background from the digitized measured data reported in 
 Fig. 3e of Ref. \cite{e2021p}.  The linear background is a straight line connecting the points for H=0T and H=1T in 
 Fig. 3. The inset on the right panel is Fig. 3c of Ref. \cite{e2021p} showing the behavior of critical field
 versus temperature extracted from the left panel, with critical field 96 mT at zero temperature and 67 mT at temperature 100K.
 }
 \label{figure1}
 \end{figure} 
 
      \begin{figure} [t]
 \resizebox{8.5cm}{!}{\includegraphics[width=6cm]{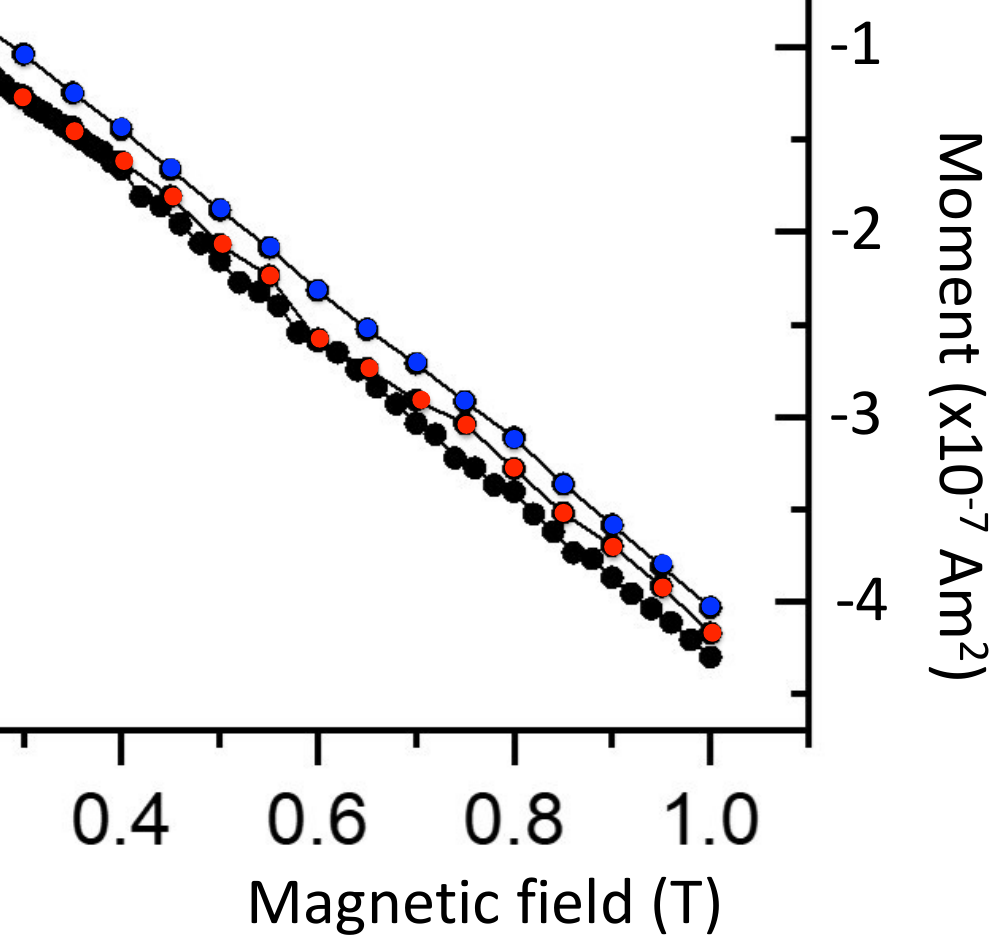}} 
 \caption {Illustration of extraction of the signals from the hysteresis loop of Fig. 10. The blue points and red points 
 are for the upper and lower branches of the hysteresis loop respectively.  
 The spacing between the colored points is 0.05T. The black points are the virgin curve, with smaller spacing.
 }
 \label{figure1}
 \end{figure} 
 
    \begin{figure*} [t]
 \resizebox{18.5cm}{!}{\includegraphics[width=6cm]{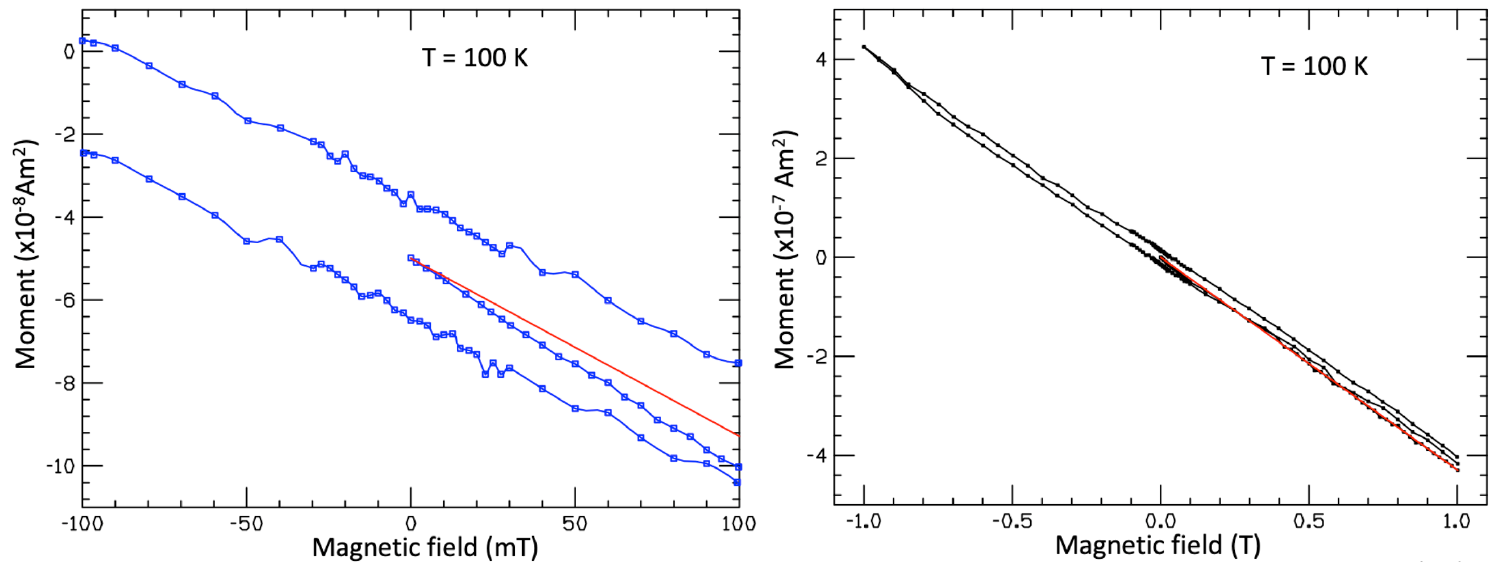}} 
 \caption {Digitization of data for T=100K shown in Figs. 3e (left panel) and Fig. S10 (right panel) of Ref. \cite{e2021p}.
 The red lines connect the points for H=0 and H=1T, with slope $-4.30\times 10^{-7} A m^2/T$.
 }
 \label{figure1}
 \end{figure*}

 \subsection{Analysis of 2022 hysteresis loop}
 
We  extracted approximate numerical data for the 100K hysteresis loop from digitization of the images shown in Fig. 3 and in 
Fig. 8 right panel, which are reproductions of Figs. 3e and S10 of Ref. \cite{e2021p}. Fig. 12 shows for illustration  the region for field $H\ge 0.3T$.  Our task was
 facilitated by the observation that both for the upper and lower branches of the hysteresis loop it can be seen that the
 points are uniformly spaced, with spacing 0.05T, in the regions $0.1T\le|H|\le1T$. This allows us to identify which are
 the virgin curve data in that range, with the exception of the range $0.1T<H<0.3T$ where the points of the 
 lower hysteresis branch overlap with the denser virgin curve points so that the distinction between the two
 is ambiguous.  The identification of the upper hysteresis branch is unambiguous over the entire field range. For field in the range $0\le|H|\le0.1T$ we obtained
 the numerical values from digitization of Fig. 8 right panel.

     \begin{figure} [t]
 \resizebox{8.5cm}{!}{\includegraphics[width=6cm]{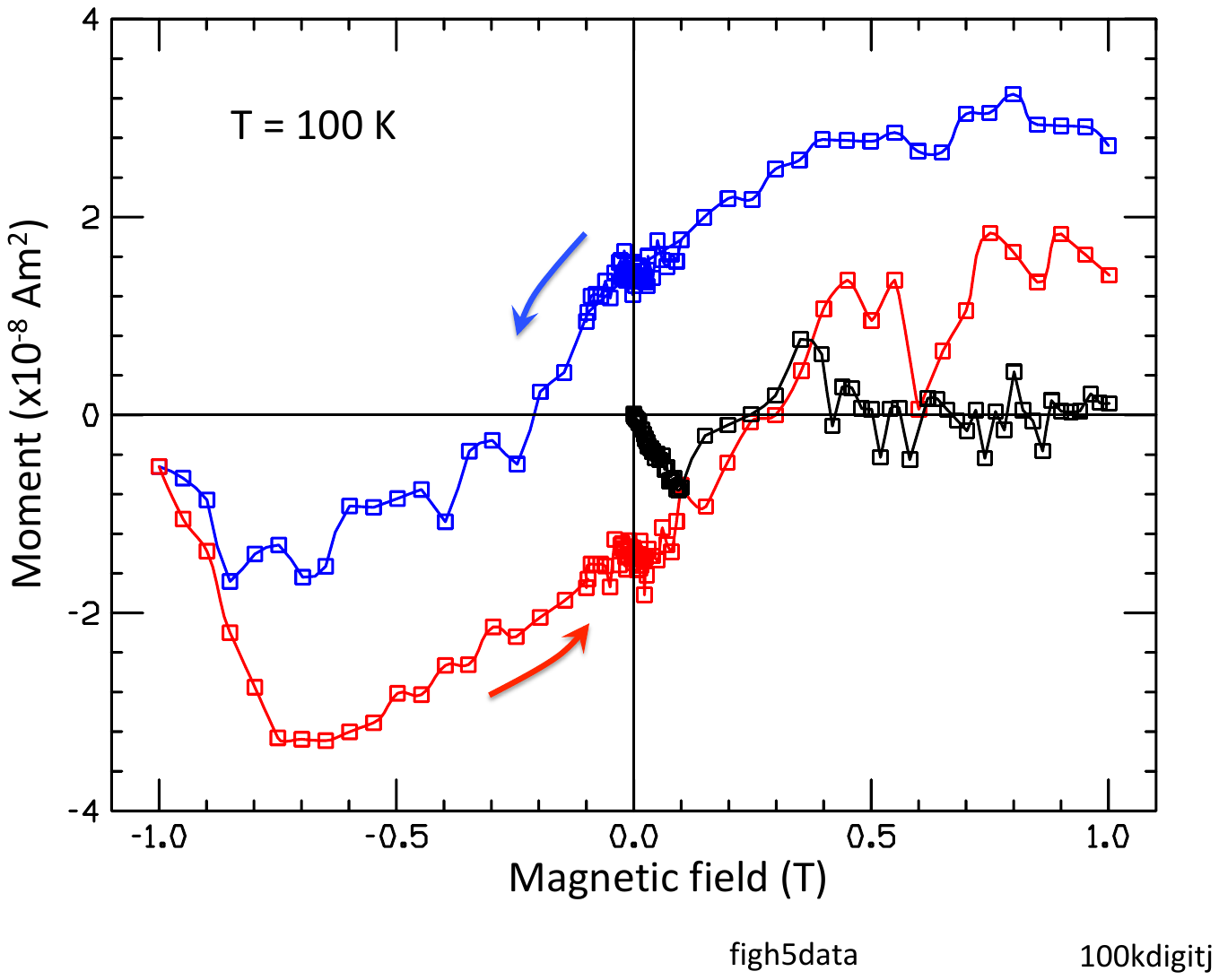}} 
 \caption {Hysteresis loop resulting from subtraction of the linear background (red line in Fig. 11) from the digitized data in Fig. 11.
 The virgin curve are the black points.
 }
 \label{figure1}
 \end{figure}

 Figure 13 shows our results. It can be seen that our digitized figures closely resemble the figures
reproduced from Ref. \cite{e2021p}, i.e. Fig. 8 right panel and Fig. 10. We have also added 
to both panels in Fig. 13 a red straight  line connecting the H=0 and H=1T points of the virgin curve. 
According to Refs. \cite{e2021p,correction}, this red line was assumed to be the background signal, and it was
 subtracted from the data to obtain the magnetic moment of the sample   shown
in Fig. 3a of Ref. \cite{e2021p}, from which estimates for the values of the lower critical magnetic
field and London penetration depth were derived \cite{e2021p}, as discussed in the previous section.


Performing the subtraction of the red line from the digitized data shown in Fig. 13, we obtain the hysteresis loop and
virgin curve shown in Fig. 14. Note that the three curves don't join at the highest field value 1T. This is surprising because
 the figure caption of Fig. S10 of Ref. \cite{e2021p} states that the field range (-1T, 1T) is 
 ``the full range of hysteresis''. Perhaps this is a misprint, and the range of fields extended to larger positive values.

 It is clear that the curves shown in Fig. 14 cannot reflect the behavior of the magnetic moment of a 
 superconducting sample. To begin with, the virgin curve crosses zero for applied field around 0.3 T.
 For a superconducting sample with upper critical field approximately 90 T, as estimated for this material,
 the magnetic moment should remain negative for fields much larger than 0.3T.  
 Furthermore, the red curve,
 denoting the magnetic moment in the return loop after having reduced the field to -1T, crosses the virgin data curve (black points)
 for magnetic field larger than 0.4T. This is impossible for a superconducting sample, the return hysteresis branch should always lie
below or coincide with the virgin curve, contrary to what is seen in Fig. 14.
Furthermore, the maximum value of the induced diamagnetic moment in the virgin curve is more than
four times smaller than the maximum moment in the loop branches, instead of being comparable to it as in all standard
superconductors.

 It may be thought that this anomalous behavior resulted from subtraction of an incorrect
 background. However, we point out that $any$ assumed background that is subtracted from the data
 in Fig. 13, assuming the background is non-hysteretic, would necessarily yield the anomalous behavior seen in Fig. 14  that the   data for the virgin curve (black points) are 
 below the data from the return loop (red points), hence
 would be incompatible with superconductivity. 
 
 To illustrate this point, we show in Fig. 15 the result obtained by assuming as background signal the 
 average of the forward and reverse M(H) field sweeps.
 This was the procedure reportedly used to obtain  hysteresis loops shown in Fig. 4a of Ref. \cite{e2021},
 \bluee{the preprint version of the published paper Ref.} \cite{e2021p}.
 It can be seen in Fig. 15 that the virgin data obtained by subtraction of this assumed background
 are again incompatible with the assumption that they reflect the behavior of a superconducting
 sample since they fall below the red curve for fields above 0.4T.
 
       \begin{figure} [t]
 \resizebox{8.5cm}{!}{\includegraphics[width=6cm]{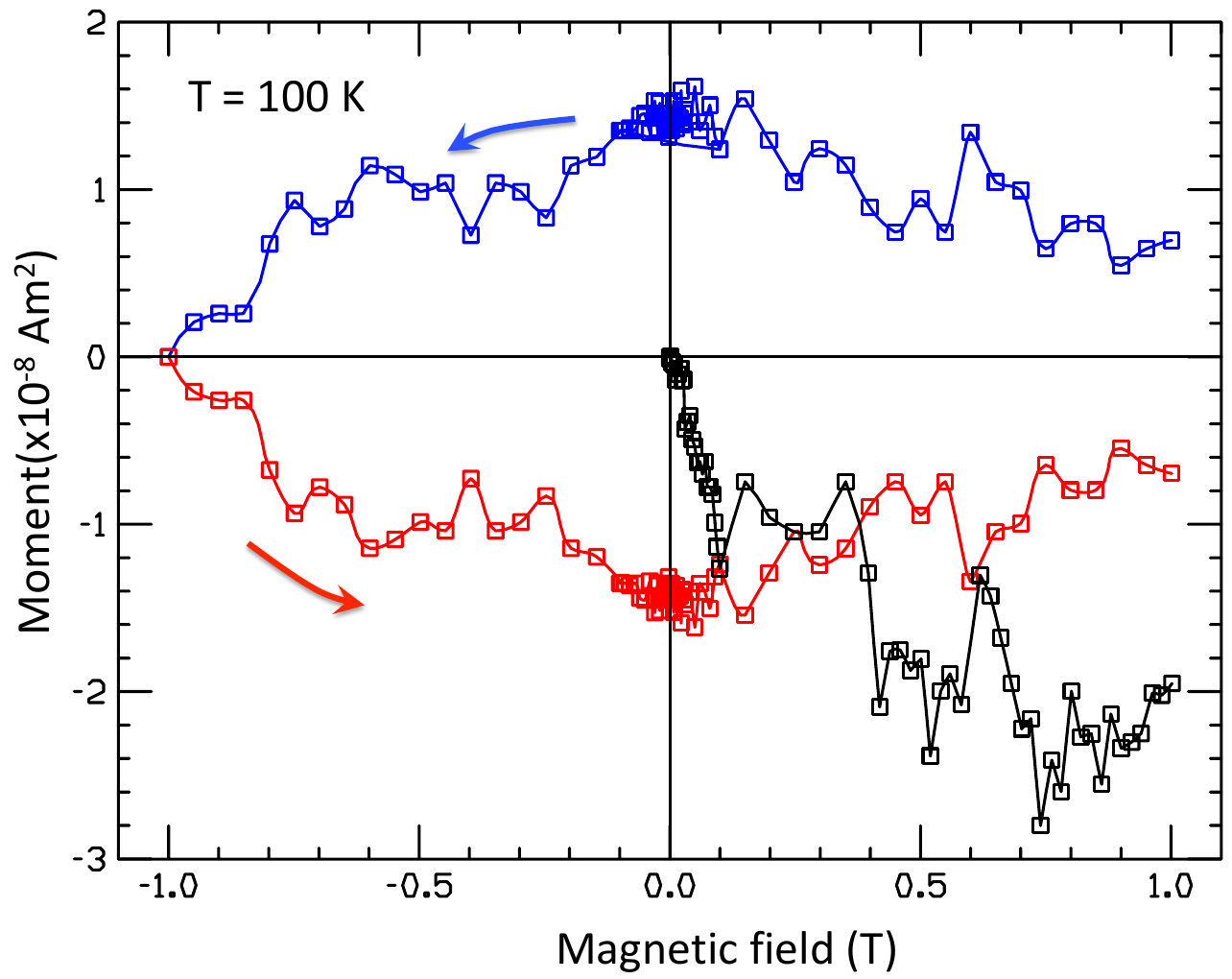}} 
 \caption {Hysteresis loop resulting from subtraction of  a background constructed from the average of the forward and reverse M(H) field sweeps as described in Ref. \cite{e2021}   from the digitized data in Fig. 4.
 The virgin curve are the black points.
 }
 \label{figure1}
 \end{figure}

 Fig. 16 shows as the green curve and points the hysteresis loop reported in 
 Fig. 4 of Ref. \cite{e2021}, obtained by subtracting from the measured loop the average of the 
 forward and reverse M(H) sweeps as explained in Ref. \cite{e2021}. It can be seen that it looks similar to the loop that we obtained
 in Fig. 15 by digitization of the data and following the same averaging and  subtraction procedure.
 We also show in Fig. 16 the magnetic moment of the sample plotted in Fig. 3a of Refs. \cite{e2021,e2021p}
 (Fig. 11 left panel here).
 As we already pointed out in Ref. \cite{hmscreening}, the fact that the moment from  Fig. 3a does not join the
 hysteresis loop in Fig. 16  indicates that  there is something very anomalous about these data.
 Now we understand the origin of this anomaly: the green loop in Fig. 16 was obtained from the
 measured data by assuming a certain background, namely the  average of the 
 forward and reverse M(H) sweeps \cite{e2021}, and the moment from Fig. 3a of Refs. \cite{e2021,e2021p}  was obtained from the measured
 virgin data by assuming a   different background, namely the straight red line in Fig. 13.   \bluee{However} it \bluee{it is not a valid procedure} to subtract different backgrounds from measured data in the same range of fields for 
 different branches of the loop.

 \subsection{Discussion}
 For reasons that have not been explained,  the miniature high-pressure cell used in the magnetic measurements of 2022 \cite{e2021p,etrapped} 
 shows a significant diamagnetic response \cite{correction}.  The authors state that they were able 
 {\it ``to obtain a pronounced diamagnetic signal from superconducting phases under high pressures”}  \cite{e2021p}.
 However the diamagnetic signal measured is the superposition of the  signal attributed to  the superconducting sample and a $much$ $larger$ diamagnetic signal
 from the background: the largest diamagnetic signal attributed to the sample is more than six times smaller than 
 the background diamagnetic signal at the same value of magnetic field. To establish that the signal of the sample is diamagnetic requires to have confidence that
 the background signal has been properly identified: if the   diamagnetic signal attributed to the background is underestimated by  $20\%$,  the resulting signal for the sample obtained by subtraction would be paramagnetic rather
 than diamagnetic. For that reason,  there has to be complete transparency on 
  what transformations were made to eliminate the background contribution from the signal,
 In our view that has not been accomplished \bluee{to date}.

        \begin{figure} [t]
 \resizebox{8.5cm}{!}{\includegraphics[width=6cm]{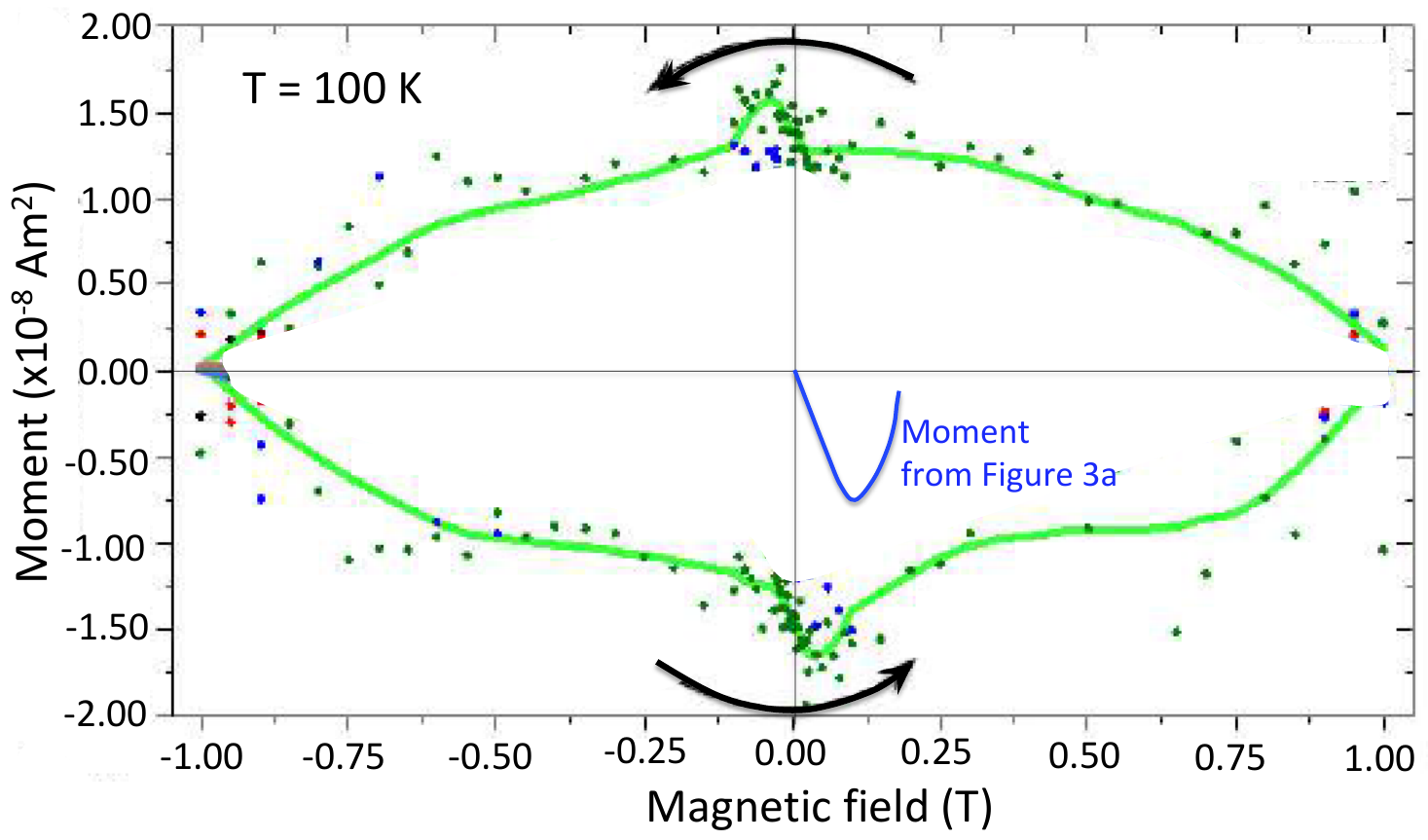}} 
 \caption {Green curve and points show the hysteresis loop reported in Fig. 4 of Ref. \cite{e2021} obtained
 according to Ref. \cite{e2021} by using as background the 
 average of the forward and reverse M(H) field sweeps.
 The blue curve shows the moment plotted in Fig. 3a of Refs. \cite{e2021,e2021p} for T=100K.
 This figure is reproduced from Ref. \cite{hmscreening}. 
 }
 \label{figure1}
 \end{figure} 
 
 In this paper we have found that the hysteresis loop and virgin curve for the sample resulting from the measurements of magnetic moment for
 sulfur hydride reported in Refs. \cite{e2021,e2021p} for temperature T=100K, under any assumption about
 the background signal, are  incompatible with the
 conclusion of Ref. \cite{e2021p} that the measurements indicate that the sample is superconducting.
 We have also found here and in Ref. \cite{bending} that the data plotted in Fig. 3a of Ref. \cite{e2021p}
 for 100K
 did not result from subtracting a linear background from the measured data, as Refs. \cite{e2021p,correction}
 states. In addition, in Ref. \cite{unpulling} we discussed anomalies in the reported data for 160K.

 It would be of great interest to determine whether these problems occur for other temperatures for which results
for magnetic moment of the sample were reported in Ref. \cite{e2021p}. In particular, no information on the measured 
hysteresis loops from which the published
results were derived is given in Ref. \cite{e2021p} for 
temperatures T = 20K,  40K, 60K, 80K. We conjecture that   the same or even more serious anomalies may be 
apparent in those  data at those lower temperatures.  None of the data underlying the measurements
reported in Ref. \cite{e2021p} have been released by the authors \bluee{to date.}  

The authors of Refs. \cite{e2022p,e2021p,etrapped} have consistently assumed that the observed 
hysteresis is only due to the sample, and that it is in fact proof that the sample is superconducting.
Experimental  support for that assumption would be provided by performing  the experiments under the same conditions of temperature
and pressure
with the non-superconducting precursor sample before the laser heating  treatment that supposedly triggers the chemical reaction that
forms the superconducting compound, and finding that no hysteresis occurs. Such  \bluee{a}   control experiment has not been reported \bluee{to date}.

Under the assumptions that the measured magnetic moments reported in Ref. \cite{e2021p} 
 are the sum of the magnetic moment of the sulfur hydride sample under study and
 a non-hysteretic background signal, we argue that our results discussed in this paper
 unequivocally rule out   that the sample is superconducting.
  If we relax the assumption that the background is non-hysteretic, the
 measured data would not rule out the possibility  that the part of the signal from the sample
originates in superconductivity, if it is assumed  that in  the M(H) return field sweep with the field increasing after the field reached large
 negative values 
 the background is in a different state (less diamagnetic)  than it was during the virgin curve measurements.   \bluee{However}  the experiment
 does  not provide evidence in favor of such \bluee{an assumption}.

If the possibility that the background is hysteretic is allowed, it is  of course no longer possible to argue that the
observed hysteresis is proof of superconductivity. It is then also  not possible to conclude that the sample is
superconducting  without additional information.

 The authors of Refs. \cite{e2022p,e2021p} have also assumed that the point where the measured magnetic moment versus field
starts to deviate from linearity is the lower critical field (corrected for demagnetization), to derive their crucial
Fig. 3c \cite{e2021p} and from it the value of the London penetration depth.
To validate that assumption it is necessary to show that for a non-superconducting sample (e.g. the precursor sample)
under the same conditions of temperature and pressure in the same DAC  the measured magnetic moment versus field  does not deviate from linearity. 
Such a control experiment has not been reported  \bluee{to date}.

 Concerning the reported measurement of trapped magnetic flux interpreted as arising from superconducting currents in Ref. \cite{etrapped},
 we argue that the results of this paper imply that the measurements of  Ref. \cite{etrapped}  provide no information on the superconductivity or 
 non-superconductivity of the samples, contrary to what is argued in Ref. \cite{etrapped}. If the background is non-hysteretic, we have shown that the measured hysteresis loop 
 cannot result from a superconducting sample. The sample could show hysteresis and trapped moments for other reasons unrelated to superconductivity \cite{hmtrapped}.
 And if the background is hysteretic, there is no way to know whether the trapped moments measured in Ref. \cite{etrapped} are due
 to the sample or the background.

\bluee{ It should also be noted that the experiments reporting trapped flux  \cite{etrapped} have not been done
under the same conditions of pressure and temperature  with a non-superconducting sample before heat treatment, to verify that
 no trapped flux results in that case, consistent with the interpretation of Ref. \cite{etrapped} that the measured
 trapped flux is due to superconductivity. Such control experiments should be done.}
 
 In summary, the results reported in this paper indicate that the experimental results reported in Refs. \cite{e2021p} and \cite{etrapped}
 are not due to superconductivity. \bluee{It cannot be ruled out that the 100K hysteresis loop analyzed here is anomalous
 and that hysteresis loops at other temperatures would show behavior consistent with the expected behavior for
 superconductors. To consider that possibility, the
 measured data for hysteresis loops for the 15 temperature values for which magnetic moment data
 were reported in Fig. 3 a and b of Ref. \cite{e2021p} should be released, so readers can analyze them and reach their
 own conclusions.}

 In conclusion we argue that our analysis here and in other papers on magnetic evidence for superconductivity in 
 hydrides under pressure \cite{persp,hmscreening,unpulling,hmtrapped,hmmre}  shows that to date  there is  no such evidence, 9 years after the onset of research in this class of materials.
 
 \section{How do you trust but verify hydride superconductivity?}
 
 {\it  I discuss the  sharp dissonance between the recent Nature Materials Editorial ``Trust but verify'' \cite{trust} and the publication, in the same journal issue,  of the Feature Comment by M. I. Eremets and coauthors on hydride superconductivity \cite{e2024}.}
 \newline

 In its January 2024 Editorial, Nature Materials urges their readers to ``Trust but verify'',  emphasizing that reproducibility of reported scientific results is key to the advancement of science.
It cites an earlier Nature Materials Editorial \cite{dataavail} that   focuses on the need for authors to share their underlying data to aid reproducibility
 and increase transparency, and decries the fact that {\it ``in the vast majority of cases data have remained locked up in the authors’ drawers, and allowed to see the light only ‘upon request’.''}

 If only. The reality is starkly different from what  the above depicts. In  reality,
 Dr. Chris Graf, Research Integrity Director of Springer Nature, the Publisher of Nature Materials, stated in a recent  letter to this author that {\it ``while we would always
encourage authors to share data with interested readers, we recognize the right of the authors to not share the data
with you''} \cite{graf}.
 
 The data in question referred to in Graf's letter are data from Minkov, Eremets and coauthors underlying  their publication on magnetic screening in hydride superconductors,
 Ref. \cite{e2021p}. I requested those data from the authors on January 11, 2023, and many times thereafter, 
 after spotting anomalies \cite{hmscreening}
 in the results published in Ref. \cite{e2021p}. The underlying data remain hidden
 to me and to the rest of the world to this date. Eremets and coauthors refuse to release them, and Nature Communications, in behavior
 endorsed by its publisher Springer Nature \cite{graf}, has declined to enforce the
 ``Data availability'' statement attached to the paper, and in addition has declined to inform readers that there are restrictions on data availability.
 
 Why are those data important? Ref. \cite{e2021p} reports how hydride materials under high pressure screen applied magnetic fields. 
 Those experimental results are of utmost importance to the field of hydride superconductivity, given the dearth of magnetic data available for those materials. However what Eremets  et al publish in Ref. \cite{e2021p} are processed data, not measured data. The measured data can potentially determine whether the materials are superconductors
 or not, adjudicating conflicting claims in the literature \cite{pers}. The published processed data \cite{e2021p}  show behavior compatible with what is expected for the magnetic behavior of superconductors. However,  because the processed data were derived from the measured data
 by a set of transformations    and {\it``linear manipulations''}, as disclosed by the authors of Ref. \cite{e2021p} in a
 recent ``Author Correction'' \cite{correction} published in response to Ref. \cite{hmscreening},  it is impossible for readers to make their own judgement about the significance of the published processed data in relation with the physical reality they purport to reflect. The authors
 of Ref. \cite{e2021p} claim for themselves the exclusive right to  know and  judge whether their data ``manipulation'' \cite{correction}    is or is not in conformance with accepted scientific practice.
 
 There is a precedent to this situation in the contentious field of hydride superconductivity. In October 2020, Ranga Dias and coauthors published
 in Nature a paper claiming room temperature superconductivity in a hydride under pressure \cite{roomt}. Immediately thereafter, having spotted anomalies
 in the paper \cite{eumine}, I requested from the authors and the journal the release of certain underlying data.
 For over a year, Ranga Dias and coauthors refused to supply those  data, claiming that  they {\it ``don’t trust Hirsch to appraise the data fairly''}  \cite{sciencenews} .
 When the underlying data were finally released \cite{diasrawdata}, more than a year after I had requested them, serious problems of incompatibility between
 published data and measured data became apparent \cite{mrecsh}, and the paper was subsequently retracted \cite{retraction1}. Those problems would not have come to light if the underlying data had not been released. The possibility that the same scenario would play out when the underlying data of
 Ref. \cite{e2021p} are released can certainly not be ruled out until those data `see the light'.
 
In fact, analysis of the data published in Ref. \cite{e2021p} has already conclusively shown \cite{unpulling,bending,hysteresis,screeningma} that  the transformations used to
go from measured to published data are $not$ exclusively linear,  contrary to what Ref. \cite{correction} claims, which invalidates the
conclusions derived from those data in Ref. \cite{e2021p}. In particular, Ref. \cite{bending}
was reviewed by 7 anonymous independent reviewers as well as the editor in Chief of the journal, some of which  independently carried out an  analysis that went beyond that contained in the paper and confirmed its findings. Eremets and coauthors were invited by the Editor of the
journal to submit a reply to Ref. \cite{bending} but have not done so.

Given this situation, it appears remarkably tone-deaf that  in the very same issue where Nature Materials encourages readers to ``Trust but verify'' \cite{trust}, it features
 an article by  Eremets and coauthors, Ref. \cite{e2024},  that claims that their experimental results on hydride superconductors are valid and reproducible, listing Ref. \cite{e2021p}
(with  its measured data kept under wraps) as supporting their claim. The Nature Materials Editorial
 advertises the Eremets featured article by stating {\it ``In contentious fields with reproducibility
issues, it is wise to see what results are on
solid ground. In a Feature, by Mikhail Eremets
and colleagues, this is done for the field
of high-pressure superconductivity, where
superhydrides formed at high pressures
(of order 100 GPa) can exhibit superconductivity
at temperatures up to 250 K.''}

This is exactly equivalent to a judge appointing the defendant as jury of their own actions. Ref. \cite{e2024} does not present any
new evidence for superconductivity in hydrides, it just  repeats contested claims. Several of the references that
Ref. \cite{e2024} cites as independently confirming hydride superconductivity have been analyzed and challenged in peer-reviewed journals \cite{pers}.
In particular, a careful analysis of measured raw data supplied by the authors of Ref. \cite{huang} and Ref. \cite{timusk} to this author
and coworkers has shown \cite{huangmine,hmtimusk}  that the relation between measured and published data raises substantial doubts on the validity of the
conclusions of Refs. \cite{huang} and \cite{timusk}. None of this is mentioned in Ref. \cite{e2024}.

In order to ``Trust but verify'', as Nature Materials urges readers to do \cite{trust}, it is indispensable that journals, research institutions
and granting agencies enforce data availability statements and regulations, particularly in situations where authors refuse to release their underlying
data invoking
loopholes in the existing regulations, such as that data requests are not a ``reasonable request'', as the authors of Ref. \cite{e2021p} have done \cite{graf}.
In such cases, the reasons the authors invoke to justify their refusal to supply the data should be scrutinized by expert  independent peer
review, and ruled invalid if appropriate, instead of accepting them at face value. That is $not$ currently being done \cite{fraud,graf},
to the detriment of  scientific progress.

\medskip
\noindent
{\bf Notes added:} 

\noindent (i) Section I  of this paper is published as \href{https://www.sciencedirect.com/science/article/pii/S0921453423001910}{Physica C 616, 1354400 (2024)}).
Section II is accepted for publication in Nat Commun (2024).
Section III  is published as \href{https://www.sciencedirect.com/science/article/pii/S09214534240001450}{Physica C 617, 1354449 (2024)}).
 Section IV is accepted for publication in Int. J. Mod. Phys. B (2024).
 
 \noindent (ii) The underlying measured data referred to in this article as unavailable have recently been made available 
 at \href{https://osf.io/7wqxb/}{https://osf.io/7wqxb/}, August 9, 2024.

  \begin{acknowledgements}
 The author is grateful to F. Marsiglio for collaboration in related work\end{acknowledgements}

 \end{document}